\documentclass[usenatbib]{mn2e}
\usepackage[pdftex]{graphicx}

\title[Optical turbulence at Dome C, Dome A and South Pole]{Mesoscale optical turbulence simulations above Dome C, Dome A and South Pole}
\author[F. Lascaux et al.]{F. Lascaux,$^{1}$\thanks{E-mail:
    lascaux@arcetri.astro.it; masciadri@arcetri.astro.it} E. Masciadri$^1$\footnotemark[1],
and S. Hagelin$^{1, 2}$ \\ $^1$INAF Osservatorio Astrofisico
di Arcetri, Largo Enrico Fermi 5, I-501 25 Florence, Italy\\
$^2$Uppsala Universitet, Department of Earth Sciences, Villav\"agen 16,
S-752 36 Uppsala, Sweden}

\begin{document}

\def\CN2{\mbox{$C_N^2$}}

\label{firstpage}
\date{Accepted 2010 ??? ??, Received 2010 ??? ??; in original form
2010 ??? ??}  
\pagerange{\pageref{firstpage}--\pageref{lastpage}}
\pubyear{2008}

\maketitle

\begin{abstract}
In two recent papers the mesoscale model Meso-NH, joint with the Astro-Meso-NH package, has been validated at Dome C, Antarctica, for the characterization of the optical turbulence. It has been shown that the meteorological parameters (temperature and wind speed, from which the optical turbulence depends on) as well as the \CN2 profiles above Dome C were correctly statistically reproduced. The three most important derived parameters that characterize the optical turbulence 
above the internal antarctic plateau: the surface layer thickness, the seeing in the free-atmosphere and in the total atmosphere showed 
to be in a very good agreement with observations. 
Validation of \CN2 has been performed using all
the measurements of the optical turbulence vertical distribution obtained in winter so far. In
this paper, in order to investigate the ability of the model to discriminate between different
turbulence conditions for site testing, we extend the study to two other potential astronomical
sites in Antarctica: Dome A and South Pole, which we expect to be characterized by different
turbulence conditions. The optical turbulence has been calculated above these two sites for
the same 15 nights studied for Dome C and a comparison between the three sites has been
performed.

\end{abstract}


\section{Introduction}

The Internal Antarctic Plateau represents a potential interesting location for astronomical applications. 
For almost a decade astronomers have shown more and more interest towards this region of the Earth thanks to its peculiar atmospheric conditions. The extreme cold temperature, the dry atmosphere, the fact that  the plateau is at more than 2500 m above the sea level, 
that the turbulence seems to develop mainly in a thin surface layer of the order of 30-40 m on the top of summits 
and that the seeing above this surface layer assumes values comparable to those obtained at mid-latitude sites get this region of the earth very appealing for astronomers. South Pole has been the first site equipped with an Observatory in the Internal Antarctic Plateau in which measurements of the optical turbulence have been done (Marks et al. 1996, Marks et al. 1999). Fifteen balloons have been launched in the winter period and it has been observed that the seeing above a surface layer of $\sim$ 220 m was very good (0.37 arcsec). Measurements of the optical turbulence at Dome C are more recent. After the first observations done in 2004 with a MASS (Lawrence et al. 2004), a series of studies done with different instrumentation have been published aiming to provide the assessment of the integrated seeing (Aristidi et al. 2005, Aristidi et al. 2009) and the vertical distribution of the optical turbulence (Trinquet et al. 2008). 

\begin{table*}
\centering
 \caption{Results obtained in Lascaux et al. (2010a) that proved the Meso-NH model reliability above Dome C in reconstructing the optical turbulence spatial distribution. Three parameters are estimated: the mean surface turbulent 
 layer (h$_{sl}$), the seeing in the free atmosphere ($\varepsilon_{FA}$) obtained integrating the \CN2 from h$_{sl}$ up to the end of the atmosphere, the total seeing ($\varepsilon_{TOT}$) obtained integrating the \CN2 from the ground up the top of the atmosphere. Beside each parameter is reported the associated standard deviation ($\sigma$) and the statistical error ($\sigma$/$\sqrt{N}$).}
 \begin{tabular}{cccccccccc}
\hline
       & h$_{sl}$    & $\sigma$    & $\sigma$/$\sqrt{N}$   &  $\varepsilon_{FA}$   &    $\sigma$    & $\sigma$/$\sqrt{N}$ & $\varepsilon_{TOT}$   &    $\sigma$    & $\sigma$/$\sqrt{N}$\\
       &       (m)      & &                            &    (arcsec)     & &                    & (arcsec)  & & \\
 \hline
Observations     & 35.3 &  19.9 & 5.1 & 0.30 & 0.70 & 0.20 & 1.60 & 0.70 & 0.20 \\  
Model  & 44.2 &  24.6 & 6.6 & 0.30 &0.67 &0.17 & 1.70 & 0.77 & 0.21 \\
  \hline
\end{tabular}
\label{tab_sum}
\end{table*}

\begin{table*}
\centering
 \caption{Geographic coordinates of Dome A, Dome C and South Pole. The altitude is in meter.}
 \begin{tabular}{|l|r|r|c|c|}
  \hline
SITE          & LATITUDE             & LONGITUDE           & MESO-NH      & MEASURED    \\
              &                      &                     & ALTITUDE (m) & ALTITUDE (m) \\
 \hline
Dome A$^*$    & 80$^{\circ}$22'00"S  &077$^{\circ}$21'11"E & 4089 & 4093 \\  
Dome C$^{**}$ & 75$^{\circ}$06'04"S  &123$^{\circ}$20'48"E & 3230 & 3233 \\
South Pole    & 90$^{\circ}$00'00"S  &000$^{\circ}$00'00"E & 2746 & 2835 \\ 
  \hline
\multicolumn{4}{|l|}{$^*$ {\scriptsize GPS measurement by Dr. X. Cui (private communication).}} \\
\multicolumn{4}{|l|}{$^{**}$ {\scriptsize GPS measurement by Prof. J. Storey (private communication).}}\\
 \end{tabular}
\label{tab0}
\end{table*}

This paper deals with a different approach to the site assessment. 
In this context we are interested in investigating the abilities of a mesoscale model (Meso-NH) in reconstructing correct optical turbulence features above different sites of the Internal Antarctic Plateau and its abilities in discriminating  the optical turbulence properties of different sites. Meso-NH (Lafore et al., 1998) is a non-hydrostatic mesoscale research model developed jointly by the Centre National des Recherches M\'et\'eorologiques (CNRM) and the Laboratoire d'A\'erologie de Toulouse, France. 
The Astro-Meso-NH package (Masciadri et al. 1999a) was first proven to be able to reconstruct realistic $\CN2$ profiles above astronomical sites by Masciadri et al. (1999b) and Masciadri et al. (2001) and statistically validated later on (Masciadri \& Jabouille, 2001, Masciadri et al. 2004, Masciadri \& Egner 2006). 
In the Astro-Meso-NH package all the main integrated astroclimatic parameters such as the isoplanatic angle, the wavefront coherence time, the scintillation rate, the spatial coherence outer scale are coded in the model (Masciadri et al. 1999a).
The model is also coded to calculate the astroclimatic parameters in finite vertical slabs (h$_{min}$, h$_{max}$) in the troposphere (Masciadri et al. 1999a, Masciadri \& Garfias 2001, Lascaux et al. 2010b). It can be therefore a useful tool for adaptive optics applications in classical as well as GLAO and/or MCAO configurations  because we can produce OT vertical distribution in whatever vertical slab we wish and with the suited vertical resolution.
More recently, Meso-NH has been statistically validated above Dome C by Lascaux et al. (2009, 2010a). The most important results obtained in these two last papers are summarized in Table \ref{tab_sum}.
Briefly, the observations at Dome C, for a set of 15 winter nights (all the available nights for which is known the optical turbulence vertical distribution), gave a mean surface layer thickness $h_{sl,obs}$ = 35.3 $\pm$ 5.1 m.
The simulated surface layer thickness obtained with the Meso-NH model ($h_{sl,mnh}$ = 44.2 $\pm$ 6.6 m) is well correlated to measurements. The statistical error is of the order of 5-6 m but the standard deviation ($\sigma$) is of the order of 20-25 m. This indicates that the statistic fluctuation of this parameter is intrinsically quite important.
The median simulated free-atmosphere seeing ($\varepsilon_{mnh,FA}$ = 0.30 $\pm$ 0.17 arcsec) as well as the median total seeing ($\varepsilon_{mnh,TOT}$ = 1.70 $\pm$ 0.21 arcsec) are well correlated to observations, respectively $\varepsilon_{obs,FA}$ = 0.3 $\pm$ 0.2 arcsec and $\varepsilon_{obs,TOT}$ = 1.6 $\pm$ 0.2 arcsec.

In the context of this paper we consider that the Meso-NH model is calibrated as shown in Lascaux et al. (2010a) i.e. it 
produces optical turbulence features in agreement with observations. We therefore apply the Meso-NH 
model with the same configuration to other two sites of the plateau: South Pole and Dome A (Table \ref{tab0}). 

\begin{table}
\caption{Meso-NH model configuration. In the second column the  horizontal resolution $\Delta$X, in the third column the number of grid points and in the fourth column the horizontal surface covered by the model domain.}
\begin{tabular}{cccc}
\hline
Domain & $\Delta$X & Grid Points & Surface \\
& (km) & & (km$\times$km) \\
\hline
Domain 1       & 25& 120$\times$120& 3000$\times$3000\\
Domain 2 & 5 &  80$\times$80&   400$\times$400 \\
Domain 3  & 1 &  80$\times$80& 80$\times$80  \\
\hline
\end{tabular}
\label{tab2}
\end{table}

Why these sites ? Dome A is an almost uncontaminated site of the plateau. It is the highest summit of the plateau and, for this reason, it is expected to be among the best astronomical sites for astronomical 
applications. The high altitude reduces the whole atmospheric path for light coming from space and above the summit the katabatic wind speed is reduced to minima values. Dome A has been proved to have the strongest thermal stability (Hagelin et al. 2008) in proximity of the ground due to the coldest temperature. Dome A is a chinese base. In the last few years the chinese astronomers gave a great impulse to the 
site characterization showing a great interest for building astronomical facilities in this site. Optical turbulence measurements during the winter time are not yet available but site testing programs are on-going (Ashley et al. 2010).
South Pole is interesting in our study because measurements of optical turbulence are available and, at the same time, the site is not located on a summit but on a gently slope. From the preliminary measurements done in the past we expect a surface turbulent layer that is thicker than the surface layer developed above the other two sites (Dome C and Dome A) due to the ground slope and the consequent katabatic winds in proximity of the surface. The three sites form therefore a perfect sample for a benchmark test on the model behavior and the model abilities.

In Section \ref{num} the numerical set-up of the model is presented. In Section \ref{opt} results of the complete analysis of the three major parameters that characterize the optical turbulence features: surface layer thickness, seeing in the free atmosphere i.e. calculated above the surface layer and total seeing are reported. Two different criteria to define the surface layer are used with consequent double treatment. Finally, in Section \ref{concl} the results of this study are summarized.

\section{Numerical set-up}
\label{num}
Meso-NH \citep{laf} can simulate the temporal evolution of the three-dimensional atmospheric flow
over any part the globe.
The prognostic variables forecasted by this model are the three cartesian
components of the wind $u$, $v$, $w$, the dry potential temperature $\Theta$, the
pressure $P$, the turbulent kinetic energy $TKE$.
\par
The system of equation is based upon an anelastic formulation
allowing for an effective filtering of acoustic waves.
A Gal-Chen and Sommerville\citet{gcs} coordinate on the vertical and a C-grid
in the formulation of Arakawa and Messinger\citet{am} for the spatial
digitalization is used.
The temporal scheme is an explicit three-time-level leap-frog scheme with a time
filter \citep{as}.
The turbulent scheme is a one-dimensional 1.5 closure scheme \citep{cux} with the
Bougeault and Lacarr\`ere\citet{bl} mixing length.
The surface exchanges are computed in an externalized surface scheme (SURFEX)
including different physical packages, among which ISBA \citep{np} for vegetation.
Masciadri et al. (1999a,b) implemented the optical turbulence package
to be able to forecast also the optical turbulence ($C_N^2$ 3D maps) and all the
astroclimatic parameters deduced from the $C_N^2$.
We will refer to the 'Astro-Meso-NH code' to indicate this package.
The integrated astroclimatic parameters
are calculated integrating the  $C_N^2$ with respect to the zenith in the
Astro-Meso-NH code.
We list here the main characteristics of the numerical configuration used in this study:
\begin{itemize}
\item The interactive grid-nesting technique \citep{st} is used, with three imbricated domains of increased horizontal mesh-sizes 
($\Delta$X=25 km, 5 km and 1 km, Table \ref{tab2}). 
Such a method is used to permit us to achieve the best resolution on a small surface but keeping the volumetric domain in which the simulation is done in thermodynamic equilibrium with the atmospherical circulation that evolves at large spatial scale on larger domains.
We shown \citep{lf09} that the simulations results are sensitive to the chosen horizontal resolution.
To achieve a good correlation between model outputs and observations, 
a grid-nesting configuration with a high horizontal resolution (at least $\Delta$X=1 km) is mandatory.
\item The vertical grid is the same for all the domains reported in Table \ref{tab2}. The first vertical grid point is at 2 m above ground level (a.g.l.). A logarithmic stretched grid 
up to 3500 m a.g.l. (with 12 points in the first hundred of meters) is employed. Above 3500 m a.g.l., the vertical resolution is constant ($\Delta$H $\sim$ 600 m). 
The maximum altitude achieved is around 20 km a.g.l.. 
The first point at only 2 m above the ground (and with 12 points in the first hundred of meters) is necessary to forecast 
the typical very thin surface layer observed in the Antartic Plateau. 
\item All simulations are initialized and forced every 6 hours at synoptic times (00:00, 06:00, 12:00, 18:00) UTC by analyses from the 
European Center for Medium-range Weather Forecasts (ECMWF)\footnote{ECMWF: $http://www.ecmwf.int/$}. 
The simulations run for 18 hours. Note that the time at which the simulation starts (UTC) differs for Dome A, Dome C and South Pole. 
This is done so to be able to compare optical turbulence profiles simulated in the same temporal interval with respect 
to the local time (LT). 
For each night, a mean vertical profile of \CN2 is computed between the time interval (20:00 - 00:00) LT as done in Lascaux et al. (2009, 
2010a). This range is centered on the time at which the balloons were typically launched at Dome C. In this way we obtain the most representative simulated \CN2 profile for each night\footnote{In the prediction of a parameterized parameter (such as the optical turbulence) 
there is not a 1-1 correlation with the real time. 
This means, with an explicative example, that is somehow meaningless to predict the turbulence at a precise time t=t$^{*}$ as we do for a 
parameter that we resolve explicitly such as the temperature or the wind speed. 
This is the reason why, in order to obtain the most representative \CN2 profile to be compared to measurements, we calculate the mean of 
the \CN2 in a temporal interval $\Delta$T. 
Such a procedure has been used in many previous papers \citep{m5,m3,m4}}. 
In Table \ref{tab1} are reported, for each site, the time at which the simulation starts and the duration $\Delta$T of the simulation with respect to the local time.  
\item An optimized version of the externalized surface scheme ISBA (Interaction Soil Biosphere Atmosphere) for antarctic 
conditions is employed \citep{lem,lem10}. 
Such a scheme has been used in Lascaux et al. (2010a) and it contributed to provide a realistic reconstruction of the optical turbulence near the surface (optical turbulence strength and turbulence layer thickness). 
It is indeed obvious that the most critical part of an atmospherical model for this kind of simulations is the scheme that controls the air/ground turbulent fluxes budget. Our ability in well reconstructing the surface temperature T$_{s}$ is related to the ability in reconstructing the sensible heat flux H that is responsible of the buoyancy-driven turbulence in the surface layer. 
\item The Astro-Meso-NH package (Masciadri et al. 1999a) implemented in the most recent version of Meso-NH has been used to calculate 
the optical turbulence and derived astroclimatic parameters. 
\end{itemize}

\begin{figure*}
\begin{center}
\includegraphics[width=\textwidth]{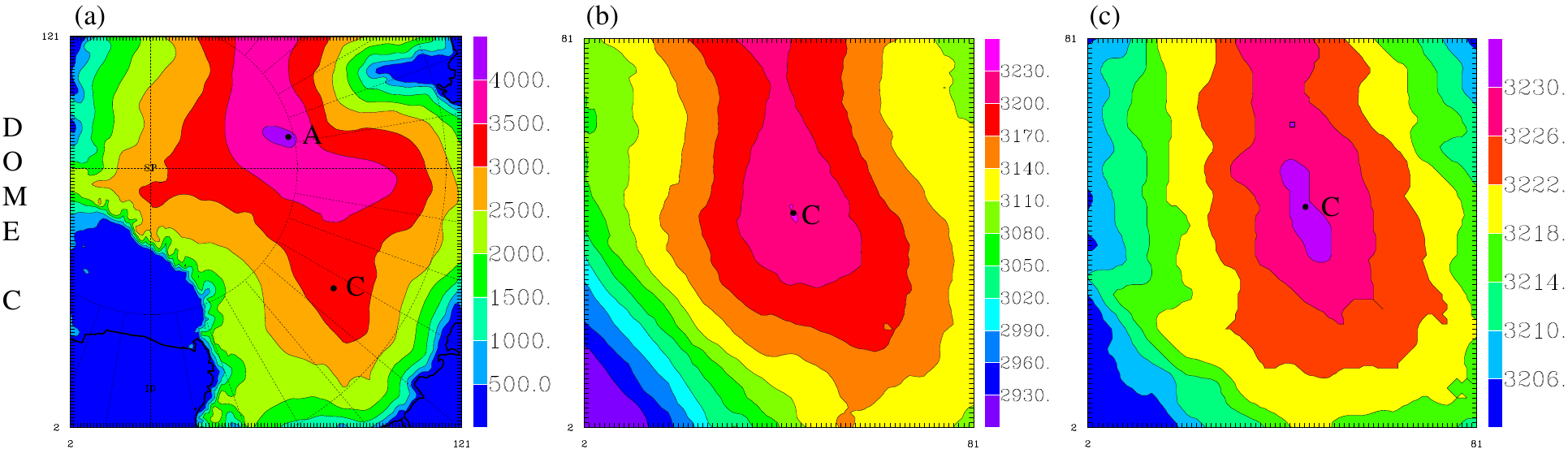}
\includegraphics[width=\textwidth]{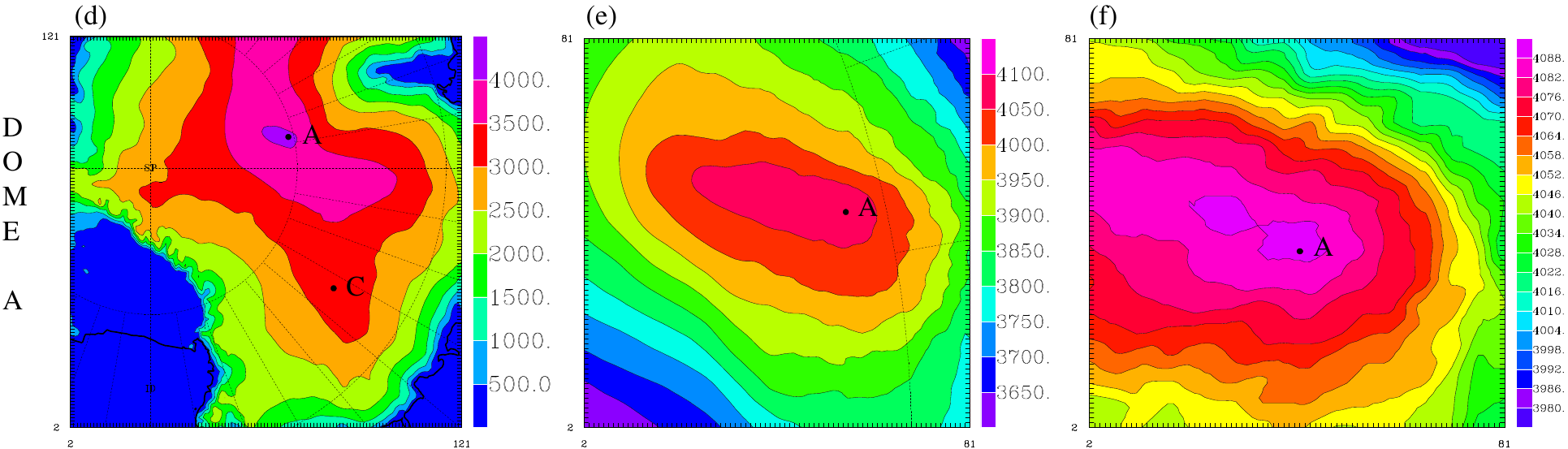}
\includegraphics[width=\textwidth]{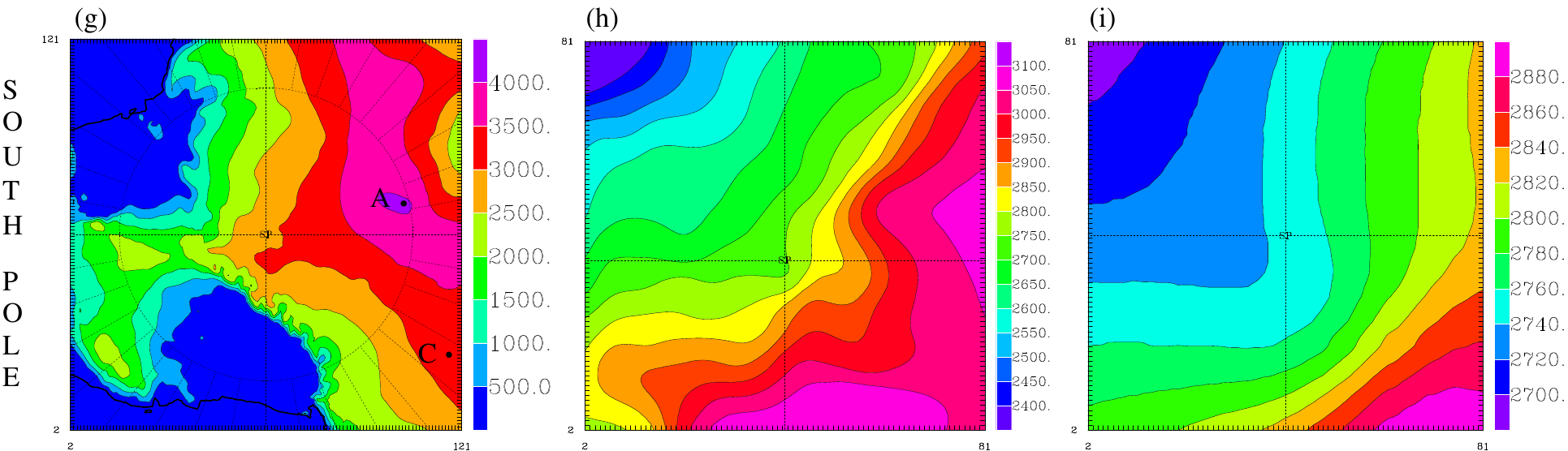}
\end{center}
\caption{Orography of three different regions of the internal Antarctic Plateau as seen by the Meson-Nh model (polar
stereographic projection, grid-nesting configuration).
(a), (b) and (c) show the three imbricated domains for the Dome C simulations, with horizontal resolution of 25 km, 5 km and 1 km, respectively.
(d), (e) and (f) show the three imbricated domains for the Dome A simulations, with horizontal resolution of 25 km, 5 km and 1 km, respectively.
(g), (h) and (i) show the three imbricated domains for the South Pole simulations, with horizontal resolution of 25 km, 5 km and 1 km, 
respectively.
The dot labeled 'C' indicates the Concordia Station. The dot labeled 'A' indicates Dome A. SP stands for South Pole. 
The altitude is expressed in meter (m).}
{\label{fig1}}
\end{figure*}

\begin{table*}
\centering
 \caption{Simulation starting time and time interval chosen for \CN2 computations for the 3 different sets of simulations 
          (Dome A, Dome C and South Pole).}
 \begin{tabular}{|c|c|c|c|}
  \hline
                    & Dome A               & Dome C               & South Pole            \\
  \hline
Starting time        & 06:00 UTC / 11:00 LT & 00:00 UTC / 08:00 LT & 12:00 UTC / 12:00 LT  \\
  \hline
Time interval        &                      &                      &                       \\
for \CN2 computations & 15:00 - 19:00 UTC    & 12:00 - 16:00 UTC    & 20:00 - 00:00 UTC     \\
(20:00 - 00:00 LT)   &                      &                      &                       \\
  \hline
 \end{tabular}
\label{tab1}
\end{table*}
As shown in \cite{lf10a}, the best choice for the description of the orography is the RAMP (Radarsat Antarctic Mapping Project) 
Digital Elevation Model (DEM) presented in \cite{liu}, instead of the GTOPO30 DEM from the U.S. Geological Survey used in \cite{lf09}.
For this study, therefore, the RAMP Digital Elevation Model has been used.  The orography of each area of interest in this study (Dome C, Dome A, South Pole) is displayed on Fig.~\ref{fig1}. All the grid-nested 
domains, from low horizontal resolution (larger mesh-size) to high horizontal resolution (smaller mesh-size) are reported.
As can be seen in Fig.~\ref{fig1} (c,f,i) the orography around Dome C and Dome A is more detailed than the orography in proximity of the South Pole. This is due to the fact that the procedure to obtain a DEM integrates data from many different sources (satellite radar altimetry, airborne surveys, 
GPS surveys, station-based radar sounding...). However the resolution of some areas (typically those that can hardly receive information from the satellites) remain poorer than others. The region included in the inner circular polar region (and therefore South Pole) fits with this condition and this is the reason why the orography is somehow less detailed than the rest of the Internal Antarctic Plateau. 
Nevertheless, this is a region with no peaks or mountains and with just a regular and gently slope. 
We can therefore reasonably expect that the poorer accuracy in the orography has little or minor influence on the results of 
the numerical simulations done with a mesoscale model such as Meso-NH.\\
The same set of 15 winter nights used by \cite{lf09,lf10a} to validate the model above Dome C is investigated in this study for the three antarctic sites 
Dome C, Dome A and South Pole. 

\section{Optical Turbulence above Dome C, Dome A and South Pole}
\label{opt}
In this section we investigate and compare the values obtained above the three sites (Dome C, Dome A and South Pole) of three parameters that characterize the optical turbulence features above the antarctic plateau:
\begin{itemize}
\item surface layer thickness;
\item free atmosphere seeing from the surface layer thickness (h$_{sl}$) up to the top of the atmosphere;
\item total seeing from the ground up to the top of the atmosphere. We note that this corresponds to $\sim$10 km because the balloons explode at this altitude due to the high pressure and the strong wind speed.
\end{itemize}
In a numerical mesoscale model the great challenge and difficulty is related to the parameterization of the optical turbulence. The critical issue is related to the ability of the model in reconstructing the vertical distribution of the optical turbulence (i.e. the $\CN2$). This is the reason why we selected and studied these three fundamental parameters. The other integrated astroclimatic parameters are obtained calculating the integral of the 
$\CN2$ and wind speed vertical profiles along the troposphere.
A forth-coming paper will be dedicated to the analysis of the integrated astroclimatic parameters.

\subsection{Optical turbulence surface layer thickness}
To compute the surface layer thickness for each night, the same method employed in \cite{tr} and \cite{lf10a} is first used. 
The thickness $h_{sl}$ is defined as the vertical slab containing 90 per cent of the optical turbulence developed inside 
the first kilometer above the ground:
\begin{equation}
\label{eq:bl1}
\frac{ \int_{8m}^{h_{sl}} C_N^2(h)dh }{ \int_{8m}^{1km} C_N^2(h)dh } < 0.90
\end{equation}
where $C_N^2$ is the refractive index structure parameter. 
We remind here that the selection of this criterium (that we call criterium A) is motivated by the fact that we intend 
to compare our calculations with measurements done by Trinquet et al. (2008). 
This criterium has been selected by Trinquet et al. (2008) because the typical optical turbulence features above the internal antarctic plateau is characterized by a major bump at the surface and a consistent decreasing of the optical turbulence strength in the first tens of meters. The selection of the percentage is obviously absolutely arbitrary and, in this context, is mainly useful to check the correlation with measurements and to compare predictions on different sites (in relative terms therefore).
The choice of the inferior limit of the integral (8 m) is motivated by the fact that Trinquet et al. (2008) intended to compare results obtained with balloons with those provided by the DIMM placed at 8 m from the ground.  

\begin{table}
\centering
\caption{Mean surface layer thicknesses $h_{sl}$ computed for the 3 sites, for 15 different winter nights using the criterion in
         Eq. \ref{eq:bl1}. Units in meter (m). The mean values are also reported  with the associated 
         statistical error $\sigma$/$\sqrt{N}$.}
\begin{tabular}{|c|r|r|r|}
\hline
Date                & Dome A & Dome C & South Pole \\
\hline
04/07/05            &  65.0  &  30.4  & 117.6      \\ 
07/07/05            & 529.4  &  35.4  & 262.9      \\
11/07/05            &  28.6  &  80.0  & 131.9      \\
18/07/05            &  27.7  &  49.7  & 224.0      \\
21/07/05            &  17.6  &  66.7  & 136.3      \\
25/07/05            &  15.7  &  27.4  & 298.6      \\
01/08/05            &  25.5  &  22.6  & 185.2      \\
08/08/05            &  53.1  &  34.2  & 104.4      \\
12/08/05            &  19.4  &  16.7  &  59.0      \\
29/08/05            &  17.4  &  91.4  & 251.3      \\ 
02/09/05            &  16.4  &  70.9  & 164.9      \\
05/09/05            & 125.2  & 338.4  & 128.0      \\  
07/09/05            &  59.8  &  52.5  & 103.6      \\
16/09/05            &  38.8  &  19.4  & 158.7      \\
21/09/05            &  20.1  &  21.0  & 148.0      \\
\hline
Mean                &  37.9* &  44.2* & 165.0      \\
\hline
$\sigma$            &  30.2* &  24.6* &  67.3      \\
\hline
$\sigma$/$\sqrt{N}$ &   8.1* &   6.6* &  17.4      \\
\hline
\multicolumn{4}{|l|}{*These values are computed without taking into account the} \\
\multicolumn{4}{|l|}{night of the 05/09/05 for Dome C and the night of 07/07/05} \\
\multicolumn{4}{|l|}{for Dome A (see text for further explanations).} \\
\end{tabular}
\label{tab3}
\end{table}

Table \ref{tab3} reports the computed values of the surface layer thickness for each night at the three sites, 
as well as the mean, the standard deviation ($\sigma$) and the statistical error ($\sigma/\sqrt N$) for the 15 nights.
For each night, the surface layer thickness is computed from a computed \CN2 profile averaged between 20 LT and 00 LT (see Table \ref{tab1} 
for hours in UT) as done in Lascaux et al. (2010a). 
The calculated mean surface layer thickness above South Pole is h$_{sl}$$=$165 m $\pm$ 17.4 m, at Dome C h$_{sl}$$=$44.2 m $\pm$ 6.6 m and 
at Dome A  h$_{sl}$$=$37.9 m $\pm$ 8.1 m. 
In this paper we are not forced anymore to use the same inferior limit of the integral in Eq. \ref{eq:bl1} (8 m) than Trinquet et al. 
(2008), and we can compute the surface layer thickness starting the integral at the ground.
Under this assumption the calculated mean surface layer thickness at South Pole 
is h$_{sl}$$=$158.7 m $\pm$ 16.2 m, at Dome C h$_{sl}$$=$45.0 m $\pm$ 7.1 m and at Dome A h$_{sl}$$=$34.9 m $\pm$ 7.9 m.
We conclude that at South Pole, h$_{sl}$ is more than three time larger than at Dome C or Dome A in both cases.
This difference is well correlated with previous observations done above South Pole. 
More precisely, observations related to 15 balloons launched during the period (20/6/1995 - 18/8/1995) indicated h$_{sl}$$=$220 m \citep{ma3}.
Measurements in that paper are done in winter but in a different year and for different nights. 
It is not surprising therefore that the matching between calculations and measurements is not perfect. 
Unfortunately the precise dates of nights studied in the paper from Marks et al. (1999) are not known. 
It is therefore not possible to provide a more careful estimate. 
It is however remarkable that the h$_{sl}$ above South Pole is substantially larger than the h$_{sl}$ above Dome C and Dome A.
Also we note that the typical thickness calculated above South Pole with a statistical sample of three months by \cite{seg} was h$_{sl}$$=$102 m.  
The authors used however a different definition of turbulent layer thickness. 
More precisely, they defined h$_{sl}$ as the elevation (starting from the lowest model level) at which the turbulent kinetic energy contains 
1 per cent of the turbulent kinetic energy of the lowest model layer. 
A comparison of this result with our calculations and with measurements is therefore meaningless. 
The same conclusion is valid for the estimates of h$_{sl}$ given at Dome C as already explained in Lascaux et al. (2009, 2010a). 
In conclusion, looking at Table \ref{tab3}, individuals values for each nights show a $h_{sl,SP}$ almost always higher than 100 m, 
with a maximum close to 300 m (2005 July 25), whereas $h_{sl,DC}$ and $h_{sl,DA}$ are always below 100 m.
Dome C and Dome A have a comparable surface layer thickness. 
For this sample of 15 nights, $h_{sl,DA}$ is 6.3 m smaller than $h_{sl,DC}$. We note also that the number of nights for which $h_{sl}$ is 
very small (inferior at 30 m) is more important at Dome A (nine instead of six at Dome C). This difference is however not really statistically reliable considering the number of the nights in the sample. For a more detailed discrimination between the h$_{sl}$ value at Dome C and Dome A we need a larger statistic. This analysis is planned for a forthcoming paper.

Looking at the results obtained night by night we can note some specific features observed in specific cases. 
Two nights (September 5 at Dome C ($h_{sl}$ = 338.4 m) and July 7 at Dome A ($h_{sl}$ = 529.4 m)) present similar characteristics: the surface layer thickness h$_{sl}$ is well larger than the observed one. In these two cases, however, as already explained in \cite{lf09,lf10a} for the case of Dome C, the large value of h$_{sl}$ does not mean that a thicker and more developed turbulence is present near the ground but it simply means that, in the first kilometer from the ground, 90 per cent of the turbulence develops in the (0, h$_{sl}$) range. 
On September 5, at Dome C, the model reconstructs the total seeing on the whole 20 km much weaker than what has been observed and more uniformly distributed and, consequently, the criterium (Eq.\ref{eq:bl1}) provides us a much larger value of h$_{sl}$. 
In both cases (on September 5 at Dome C and on July 7 at Dome A),  when we look at the vertical distribution of the \CN2 calculated by the model, we observe that the turbulence is concentrated well below 20 m in a very thin surface layer with a very weak total seeing (see next section).
The case of 5 September at Dome A, is however a case in which the model reconstructed a surface turbulent layer thicker than what has been observed.

It is known that mesoscale model provide a temporal variability of the turbulence in the high part of the atmosphere that is smoother 
than what observed with vertical profiler. 
This is due to the fact that a mesoscale model is more active in the low part of the atmosphere where the orographic effects are mainly 
present. 
We recently obtained \citep{lf09} very encouraging results showing that the \CN2 in the free atmosphere has a temporal variability even on 
a small dynamic range (-18, -16.5 in logarithmic scale). 
This is a signature of the improvement of the model activity in the high part of the atmosphere. 
At present, however, it presents a hazard
to quantifying the typical time-scale for temporal variability of all
the parameters related to the optical turbulence reconstructed by a
mesoscale model.
Nevertheless we can describe the temporal variability of the morphology of these parameters such as, for example the thickness of the surface layer. 

In Appendix A, we report the temporal evolution of the calculated \CN2 for all the nights above the three sites.
Looking at these pictures we can give a description of the morphology of the temporal variation of the surface layer.
Above Dome C and Dome A, the thickness of the surface layer remains mostly stable during the night, even though we have night to night 
variations as shown in Table \ref{tab3}. This fits with preliminary results shown in Ashley et al. (2010) above Dome A.
Above South Pole, the thickness varies in a much more important way during the night with oscillations that can reach 50 to 100 m. The larger variability of the typical turbulent surface layer thickness is confirmed also by the larger value of $\sigma$ observed above South Pole (Table \ref{tab3}, \ref{tab_TKE}, \ref{tab_TKEbis}). 

In order to compare our calculations and results with those obtained by \cite{seg} we applied also a different criterium (criterium B) based on the analysis of the  vertical profile of turbulent kinetic energy (TKE) instead of the vertical profile of \CN2. The TKE is certainly an ingredient from which the optical turbulence depends on and it represents the dynamic turbulent energy. 
However, it is known (Masciadri \& Jabouille 2001) that the $\CN2$ depends also on the gradient of the potential temperature and moreover, 
the selection of the value of percentage of the turbulent kinetic energy (1$\%$, 10$\%$, other...) used as a threshold is absolutely 
arbitrary.
This method is therefore not useful to quantify the absolute value of h$_{sl}$ to be compared to measurements provided by 
Trinquet et al. (2008) and Marks et al. (1999). 
It can possibly be useful for relative comparisons between different sites or to compare our calculations with calculations 
provided by Swain \& Gallee (2006).

Using this method (Table \ref{tab_TKE}), the surface layer height is determined as the elevation at which the TKE is X\% of the lowest elevation value. We calculated the h$_{sl}$ for X = 1 (Table \ref{tab_TKE}) and X = 10 (Table \ref{tab_TKEbis}). X=1 is the case treated by \cite{seg}. For each simulation, we first compute the average of the TKE profile for the night between 20 LT and 00 LT. While the average of the $\CN2$ profile is calculated with a 2 minutes rate sample, the average of the TKE is calculated with 5 profiles, available at each hour (20, 21, 22, 23, 00) LT.  This gives us an averaged vertical profile of TKE characteristic of the considered night. 

The computation of the surface layer thickness is then performed using this averaged TKE profile.
It has been observed that, when the night presents only low dynamic turbulence (with a very low averaged TKE at the 
lowest elevation level), it is very hard to retrieve a surface layer height using this criterium. This means that the turbulence is so weak that we are at the limit of necessary turbulent kinetic energy to resolve the turbulence itself.
For these nights (indicated with an asterisk in Table \ref{tab_TKE}) it could happen that we calculated the average on a number of estimates smaller than 5 (as for all the other cases). The results are reported in Table \ref{tab_TKE}.

\begin{table}\
\centering
\caption{Mean surface layer thicknesses $h_{sl}$ computed for the 3 sites, for the same set of nights shown in Table \ref{tab3}, but 
         computed with a different criterion. The surface layer height is determined as the elevation at which the averaged TKE 
         between 20 LT and 00 LT for each nigh is 1\% of the averaged lowest elevation value. Units in meter (m). The mean values are also 
         reported  with the associated statistical error $\sigma$/$\sqrt{N}$.}
\begin{tabular}{|c|r|r|r|}
\hline
Date                & Dome A & Dome C & South Pole \\
\hline
04/07/05            &    78* &    32  &   112      \\
07/07/05            &    6* &    32  &   112*     \\
11/07/05            &    40  &    76  &   174      \\
18/07/05            &    32  &    48  &   242      \\
21/07/05            &    22  &    56  &   144      \\
25/07/05            &    22  &    12* &   148      \\
01/08/05            &    32  &    22  &   186      \\
08/08/05            &    56  &    32  &   112      \\
12/08/05            &    22  &    60  &    58      \\
29/08/05            &    22  &    82  &   250      \\
02/09/05            &    20  &    60  &   188      \\
05/09/05            &   136  &    30* &   146      \\
07/09/05            &    72  &    74  &   250      \\
16/09/05            &    56  &    22  &   192      \\
21/09/05            &    26  &    22  &   170      \\
\hline
Mean                &    42.8&    44&   165.6    \\
\hline
$\sigma$            &    33.0&    22.6  &    55.5    \\
\hline
$\sigma$/$\sqrt{N}$ &     8.5&     5.8&    14.3    \\
\hline
\multicolumn{4}{|l|}{*These values are computed using a number } \\
\multicolumn{4}{|l|}{of profiles minor than five.} \\
\end{tabular}
\label{tab_TKE}
\end{table}

Table \ref{tab_TKE} shows that results obtained with the criterium of the TKE are similar to those obtained with the criterium described in Eq.\ref{eq:bl1}. 
Table \ref{tab_TKEbis} provides smaller values of h$_{sl}$ above all the three sites. We treat the case (X = 10) to show that, tuning the value of the percentage, it is possible to find different values of h$_{sl}$. This means that h$_{sl}$ estimates are useful only if they are compared 
to measurements using the same criteria. 

\begin{table}
\centering
\caption{Mean surface layer thicknesses $h_{sl}$ computed for the 3 sites, for the same set of nights shown in Table \ref{tab3}, but 
         computed with a different criterion. The surface layer height is determined as the elevation at which the averaged TKE 
         between 20 LT and 00 LT for each nigh is 10\% of the averaged lowest elevation value. Units in meter (m). The mean values are also 
         reported  with the associated statistical error $\sigma$/$\sqrt{N}$.}
\begin{tabular}{|c|r|r|r|}
\hline
Date                & Dome A & Dome C & South Pole \\
\hline
04/07/05            &    58 &    22  &   56      \\
07/07/05            &    2 &    24  &   62     \\
11/07/05            &    28  &    52  &   64      \\
18/07/05            &    22  &    32  &   186      \\
21/07/05            &    14  &    40  &   110     \\
25/07/05            &    12  &    6 &   102     \\
01/08/05            &    22  &    16  &   126      \\
08/08/05            &    40  &    24  &   102      \\
12/08/05            &    16  &    8  &    40      \\
29/08/05            &    14  &    68  &   192      \\
02/09/05            &    14  &    46  &   112      \\
05/09/05            &   106  &    30 &   88      \\
07/09/05            &    50  &    52  &   68      \\
16/09/05            &    40  &    12  &   150      \\
21/09/05            &    18  &    14  &   116     \\
\hline
Mean                &    30.4 &    27&   104.9    \\
\hline
$\sigma$            &    26.1&    15.6  &    45.2    \\
\hline
$\sigma$/$\sqrt{N}$ &     6.7&     4&    11.7    \\
\hline
\end{tabular}
\label{tab_TKEbis}
\end{table}

To conclude, both criteria (A and B with X=1) give similar mean $h_{sl}$ values for all the 3 sites for this limited set of nights.
Evaluating the surface layer thickness over a more extended set of nights should be the next step. It would permit us to compute more 
reliable and robust statistical estimates for $h_{sl}$ over the 3 antarctic sites and possibly to identify discrimination between the $h_{sl}$  at 
Dome A and Dome C. 
This also means that our estimate of h$_{sl}$=165 m above South Pole is better correlated to measurements (h$_{sl}$=220 m) 
than the estimate (h$_{sl}$=102 m) obtained by \cite{seg} at the same site. 

\subsection{Seeing in the free atmosphere and seeing in the whole atmosphere}
The seeing in the free atmosphere and in the whole atmosphere for $\lambda$$=$0.5$\times$10$^{-6}$m is:
\begin{equation}
\varepsilon_{FA}=5.41 \cdot \lambda^{-1/5} \cdot \left( \int_{h_{sl}}^{h_{top}} C_N^2(h) \cdot dh \right) ^{3/5}
\end{equation}
\begin{equation}
\varepsilon_{TOT}=5.41 \cdot  \lambda^{-1/5} \cdot \left( \int_{8m}^{h_{top}} C_N^2(h) \cdot dh \right) ^{3/5}
\end{equation}
with h$_{top}$ $\sim$ 13 km from the sea level i.e. where the balloons explode and we have no more their signal.
Table \ref{tab4} shows the simulated total seeing ($\varepsilon_{TOT}$) and free-atmosphere seeing ($\varepsilon_{FA}$) for 
each night and each sites (Dome C, Dome A and South Pole). 
We define the free atmosphere as the portion of the atmosphere extended from the mean $h_{sl}$ reported in Table \ref{tab3} up to $h_{top}$. 
The median values of the seeing as well as the standard deviation ($\sigma$) and the statistical error ($\sigma/\sqrt N$) are reported.
As expected the total seeing is stronger at Dome A ($\varepsilon_{TOT,DA}$ = 2.37 $\pm$ 0.27 arcsec) than at Dome C 
($\varepsilon_{TOT,DC}$ = 1.70 $\pm$ 0.21 arcsec) or South Pole ($\varepsilon_{TOT,SP}$ = 1.82 $\pm$ 0.23). 
The total seeing is very well correlated with measurements at Dome C (Lascaux et al. 2010a - $\varepsilon_{TOT,obs}$ = 1.6 arcsec) and at South Pole (Marks et al. 1999 - $\varepsilon_{TOT,obs}$ = 1.86 arsec) getting the estimate at Dome A highly reliable.
The minimum median free-atmosphere seeing is found at Dome A ($\varepsilon_{FA,DA}$ = 0.23 $\pm$ 0.28 arcsec). The medain free-atmosphere seeing at Dome C is $\varepsilon_{FA,DC}$ = 0.30 $\pm$ 0.17 arcsec and at South Pole, $\varepsilon_{FA,SP}$ = 0.36 $\pm$ 0.11 arcsec. The seeing in the free atmosphere is very well correlated with measurements at Dome C (Lascaux et al. 2010a - $\varepsilon_{FA,obs}$ = 0.30 arcsec) and at South Pole (Marks et al. 1999 - $\varepsilon_{FA,obs}$ = 0.37 arsec) getting again very reliable the method (Meso-NH model) as well as the estimates at Dome A.
What is remarkable is that, even if $h_{sl,DA}$ $<$ $h_{sl,DC}$ $<$ $h_{sl,SP}$, Dome A is the site with the lowest free-atmosphere seeing $\varepsilon_{FA}$. This means that at Dome A as well as at Dome C the turbulence is concentrated inside the first tens of meters 
from the ground. Moreover, the turbulence in the surface layer is stronger at Dome A than at Dome C. This can be explained with the 
stronger thermal stability of Dome A near the ground. Our results match, therefore, with predictions we did in Hagelin et al. (2008) studying 
only features of the meteorological parameters.

\begin{table}
\caption{Total seeing $\varepsilon_{TOT}$$=$$\varepsilon_{[8m,h_{top}]}$ and seeing in the free atmosphere
$\varepsilon_{FA}$$=$$\varepsilon_{[h_{sl},h_{top}]}$ calculated for the 15 nights and averaged in the temporal range
20-00 LT.
See the text for the definition of h$_{sl}$ and h$_{top}$.}
\begin{tabular}{cccc}
\hline
& DOME A      & DOME C               & SOUTH POLE         \\
\hline
 Date     & $\varepsilon_{FA}$/$\varepsilon_{TOT}$      & $\varepsilon_{FA}$/$\varepsilon_{TOT}$      & $\varepsilon_{FA}$/
$\varepsilon_{TOT}$        \\
          & {\tiny($h_{sl}$=37.9m)} & {\tiny($h_{sl}$=44.2m)} & {\tiny($h_{sl}$=165m)} \\
\hline
 04/07/05 &      2.55 / 3.37      &     0.22 / 2.28      &    0.40 / 1.67         \\
 07/07/05 &      0.20 / 0.24      &     0.28 / 1.91      &    0.31 / 0.70         \\
 11/07/05 &      0.23 / 2.78      &     1.61 / 1.81      &    0.47 / 1.96         \\
 18/07/05 &      0.21 / 2.73      &     0.80 / 1.94      &    1.46 / 2.28         \\
 21/07/05 &      0.21 / 1.95      &     0.86 / 1.27      &    0.31 / 1.71         \\
 25/07/05 &      0.22 / 1.55      &     0.25 / 0.85      &    0.32 / 0.76         \\
 01/08/05 &      0.22 / 1.78      &     0.22 / 2.27      &    0.52 / 1.78         \\
 08/08/05 &      1.45 / 2.42      &     0.35 / 1.70      &    0.28 / 1.69         \\
 12/08/05 &      0.23 / 2.37      &     0.23 / 0.99      &    0.29 / 1.82         \\
 29/08/05 &      0.23 / 1.83      &     2.29 / 2.47      &    1.55 / 2.11         \\
 02/09/05 &      0.22 / 1.76      &     1.16 / 1.54      &    0.81 / 3.56         \\
 05/09/05 &      3.21 / 3.36      &     0.30 / 0.52      &    0.31 / 2.98         \\
 07/09/05 &      2.43 / 3.49      &     1.69 / 3.73      &    0.31 / 1.41         \\
 16/09/05 &      1.11 / 4.60      &     0.21 / 1.57      &    0.99 / 3.96         \\
 21/09/05 &      0.20 / 2.30      &     0.26 / 1.63      &    0.36 / 2.32         \\
\hline
 Median   &      0.23 / 2.37      &     0.30 / 1.70      &    0.36 / 1.82         \\
\hline
 $\sigma$ &      1.08 / 1.03      &     0.67 / 0.77      &    0.43 / 0.90         \\
\hline
 $\sigma$/$\sqrt{N}$ & 0.28 / 0.27&     0.17 / 0.21      &    0.11 / 0.23         \\
\hline
\end{tabular}
\label{tab4}
\end{table}

At South Pole, however, the \CN2 vertical distribution decreases in a less abrupt way because
the thermal stability near the ground is less important.
The \CN2 vertical distribution is spread over hundreds of meters from the ground,
instead of tens of meters like for Dome A or Dome C. As a consequence the total seeing is also 
weaker than above Dome C and Dome A.

\begin{figure*}
\begin{center}
\includegraphics[width=\textwidth]{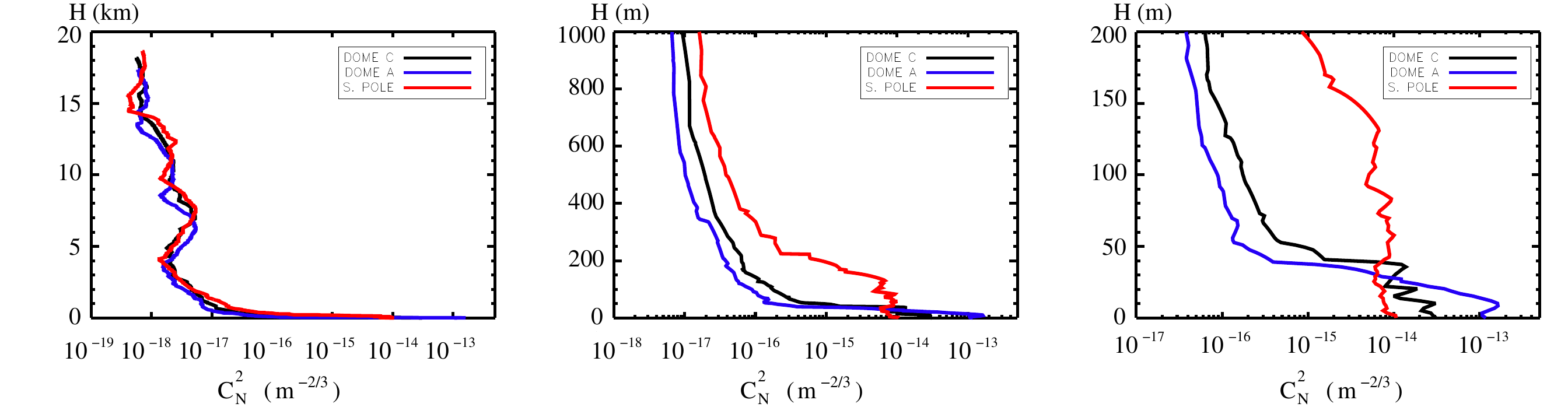}
\end{center}
\caption{Median \CN2 profiles simulated with the Meso-NH mesoscale model at Dome C 
(black), Dome A (blue) and South Pole (red). Left: from the ground up to 20 km. Middle: 
from the ground up to 1 km. Right: from the ground up to 200 m. Units are $m^{-2/3}$.}
\label{fig2}
\end{figure*}

Such a behavior is evidenced in Figure \ref{fig2}, which displays the median vertical
\CN2 profiles over the 3 sites.

Looking at Table \ref{tab4}, we note that the values of $\sigma$ for the total seeing above the three sites is
mostly comparable with no significant differences even if Dome C seems a little smaller (0.77) than Dome A (1.03) and South Pole (0.90). This indicates a comparable variability of the turbulence above the three sites. For what concerns
the seeing in the free atmosphere, the value of $\sigma$ above Dome A is almost double (1.08) than above
Dome C (0.67) and South Pole (0.43).

\begin{table*}
\centering
 \caption{Summary of the main results obtained in this study: surface layer h$_{sl}$, seeing in the free atmosphere 
($\varepsilon_{FA}$) and total seeing ($\varepsilon_{TOT}$) at Dome C, Dome A and South Pole. 
The associated standard deviation ($\sigma$) and the statistical error ($\sigma$/$\sqrt{N}$) are also reported.}
 \begin{tabular}{cccccccccc}
  \hline
       & h$_{sl}$    & $\sigma$    & $\sigma$/$\sqrt{N}$   &  $\varepsilon_{FA}$   &    $\sigma$    & $\sigma$/$\sqrt{N}$ & $\varepsilon_{TOT
}$   &    $\sigma$    & $\sigma$/$\sqrt{N}$\\
       &       (m)      & &                            &    (arcsec)     & &                    & (arcsec)  & & \\
 \hline
Observations - Dome C    & 35.3 &  19.9 & 5.1 & 0.30 & 0.70 & 0.20 & 1.60 & 0.70 & 0.20 \\
Meso-NH - Dome C         & 44.2 &  24.6 & 6.6 & 0.30 & 0.67 & 0.17 & 1.70 & 0.77 & 0.21 \\
Meso-NH - Dome A         & 37.9 &  30.2 & 8.1 & 0.23 & 1.08 & 0.28 & 2.37 & 1.03 & 0.27 \\   
Meso-NH - South Pole     &165.0 &  67.3 & 17.4& 0.36 & 0.43 & 0.11 & 1.82 & 0.90 & 0.23 \\
  \hline
\end{tabular}
\label{tab_sum2}
\end{table*}

\section{Conclusion}
\label{concl}
In this study the mesoscale model Meso-NH was used to perform forecasts of optical turbulence (evolutions of \CN2 profiles) 
for 15 winter nights at three different antarctic sites: Dome A, Dome C and South Pole. The model has been used with the same configuration 
previously validated at Dome C (Lascaux et al. 2010a) and simulations of the same 15 nights have been performed above the three sites. 
The idea behind our approach is that once validated above Dome C, the model can be used above two other sites of the internal antarctic 
plateau to discriminate optical turbulence features typical of other sites. This should show the potentiality of the numerical tool in the context of the site selection and characterization in astronomy.  South Pole has been chosen because in the past some measurements of the optical turbulence have been done and this can represent a useful constraint for the model itself. For Dome A there are not at present time measurements of the optical turbulence and this study provides therefore the first estimates ever done of the optical turbulence above this site. 
We test this approach above the antarctic plateau because this region is particularly simple from the topographic point of view and certainly simpler than typical mid-latitude astronomical sites. No major mountain chains are present and the local surface circulations is mainly addressed by the energy budget air/ground transfer, the polar vortex circulation at synoptic scale and the katabatic winds generated by gravity effects on gently slopes due to the cold temperature of the iced surface. 
The main results we obtained are summarized in Table \ref{tab_sum2} and listed here:
\begin{itemize}
\item We provide the first estimate of the optical turbulence extended on the whole 20 km above the Internal Antarctic Plateau. 
\item The Meso-NH model achieves to reconstruct the three most important parameters used to characterize the optical turbulence: the turbulent surface layer thickness, the seeing in the free atmosphere and in the surface layer for the three selected sites: Dome C, Dome A and South Pole showing results in agreement with expectations. Measurements taken at Dome C and South Pole corresponds to balloons launched during 15 nights, in both cases. The statistic is not very large but reliable for a first significant result. The selected nights correspond to the 15 nights for which measurements of the Dome C are available. 
\item Dome C and Dome A present a very thin surface layer size ($h_{sl,DA}$ = 37.9 $\pm$ 8.1 m and  $h_{sl,DC}$ = 44.2 $\pm$ 6.6 m) while South Pole surface layer is much thicker ($h_{sl,SP}$ = 165 $\pm$ 17.4 m). If we apply the criterium (A) described by Eq.(1) integrating from the ground instead than 8 m from the ground we find similar result within a couple of meters.
All these estimates are well correlated with measurements. 
Surface layers calculated by the model at Dome C and Dome A have a comparable thickness considering the actual sample.
To better discriminate between the Dome A and Dome C surface layer thickness a richer statistic is necessary. 
An on-going study has started addressing this issue.
\item Dome A is the site with the strongest total seeing (2.37 $\pm$ 0.27 arcsec) with respect to Dome C ($\varepsilon_{TOT,DC}$ = 1.70 $\pm$ 0.21 arcsec) and South Pole ($\varepsilon_{TOT,SP}$ = 1.82 $\pm$ 0.23 arcsec). This is explained by the stronger thermal stability near the ground with respect to the other two sites that cause large values of the optical turbulence in the thin surface layer.
\item All the three sites show a very weak seeing in the free atmosphere i.e. above the correspondent mean h$_{sl}$: $\varepsilon_{FA,DA}$ = 0.23 $\pm$ 0.28 arcsec at Dome A,  $\varepsilon_{FA,DC}$ = 0.30 $\pm$ 0.17 arcsec at Dome C and $\varepsilon_{FA,SP}$ = 0.36 $\pm$ 0.11 arcsec at South Pole. Dome A show the weakest seeing in the free atmosphere. 
\item The temporal variability of the thickness of the surface layer is more important at South Pole than above Dome A and Dome C that show a very stable trends in agreement with observations. The temporal variability of the seeing in the whole atmosphere does not show important differences above the three sites, while the variability of the seeing in the free atmosphere is almost double at Dome A than at Dome C and South Pole.
\item Both, the total seeing and the seeing in the free atmosphere calculated by Meso-NH, are very well correlated with measurements at Dome C and South Pole getting the predictions done at Dome A highly reliable.
\item Dealing with the criteria used to define the surface layer thickness, we proved that, at least on the sample of 15 nights investigated, the criterium defined by Eq.\ref{eq:bl1} (criterium A) and the criterium using the vertical profile of the turbulent kinetic energy (TKE) taking h$_{sl}$ as the height at which the value of the TKE is less than 1$\%$ of the TKE at the lowest level near the ground (criterium B) provide very similar results. 
\item The mean h$_{sl}$ we estimate at Dome C (h$_{sl}$=44.2) is slightly thicker than what found by \cite{seg} (h$_{sl}$=27.7 m) with comparable discrepancy from measurements (h$_{sl}$ = 35.3 $\pm$ 5.1 m). The h$_{sl}$ we estimate at South Pole ($h_{sl,SP}$ = 165 $\pm$ 17.4 m) is thicker than what estimated by \cite{seg} ($h_{sl,SP}$ = 102) but better correlated to measurements ($h_{sl,SP}$ = 220 m) than what found by \cite{seg}. The h$_{sl}$ we estimate at Dome A ($h_{sl,DA}$ = 37.9 $\pm$ 8.1 m) is somehow thicker than what estimated by \cite{seg} ($h_{sl,DA}$ = 18 m). It is however important to note that the standard deviation of h$_{sl}$ is of the order of h$_{sl}$ itself or even larger. The statistic error $\sigma$/$\sqrt(N)$ is of the order of $\sim$ 10 m. We think therefore that at present there are no major differences in our results with respect to \cite{seg} with exeption of the fact that we proved that, with our model, the horizontal resolution of 1 km provides better results than a resolution of 100 km that is used by \cite{seg}.
\end{itemize}
All these results deserve now a confirmation provided by an analysis done with a richer statistical sample. Also it would be interesting to refine this study when OT measurements above Dome A will be published. Besides, we can state that all major expectations concerning the typical features of the optical turbulence above South Pole, Dome C and Dome A have been confirmed by this study. 
The tendency shown by the model is obviously that in summer time, in proximity of the surface, due to the less stable regime, the turbulence thickness increases but the turbulence strength decreases. This is however, out the goals of this paper.

\section*{Acknowledgments}

ECMWF products are extracted from the catalogue MARS,
http://www.ecmwf.int, access to these data was authorized by the
Meteorologic Service of the Italian Air Force. This study has been
funded by the Marie Curie Excellence Grant (FOROT) -
MEXT-CT-2005-023878.

\appendix

\section[]{Computed temporal evolutions of \CN2 vertical profiles for each nights at Dome C, Dome A and South Pole}
\begin{figure*}
\begin{center}
\includegraphics[width=\textwidth]{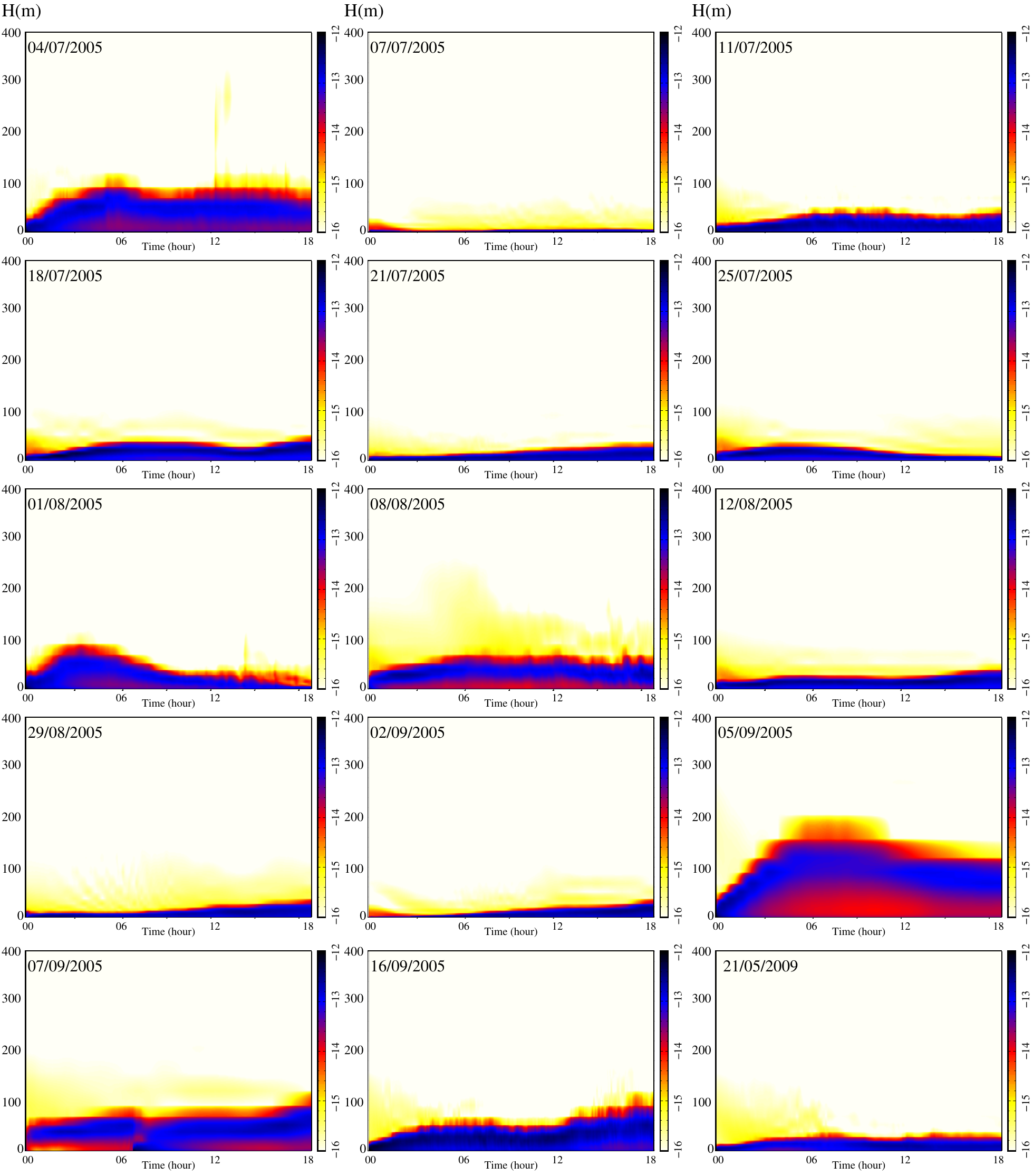}
\end{center}
\caption{Meso-NH temporal evolutions of \CN2 vertical profiles at Dome A (log units) for the 15 forecasted nights, 
for all the 18 hours of the simulations, from the ground up to 400 m above ground level. Units are $m^{-2/3}$.}
\label{fig1_app}
\end{figure*}

\begin{figure*}
\begin{center}
\includegraphics[width=\textwidth]{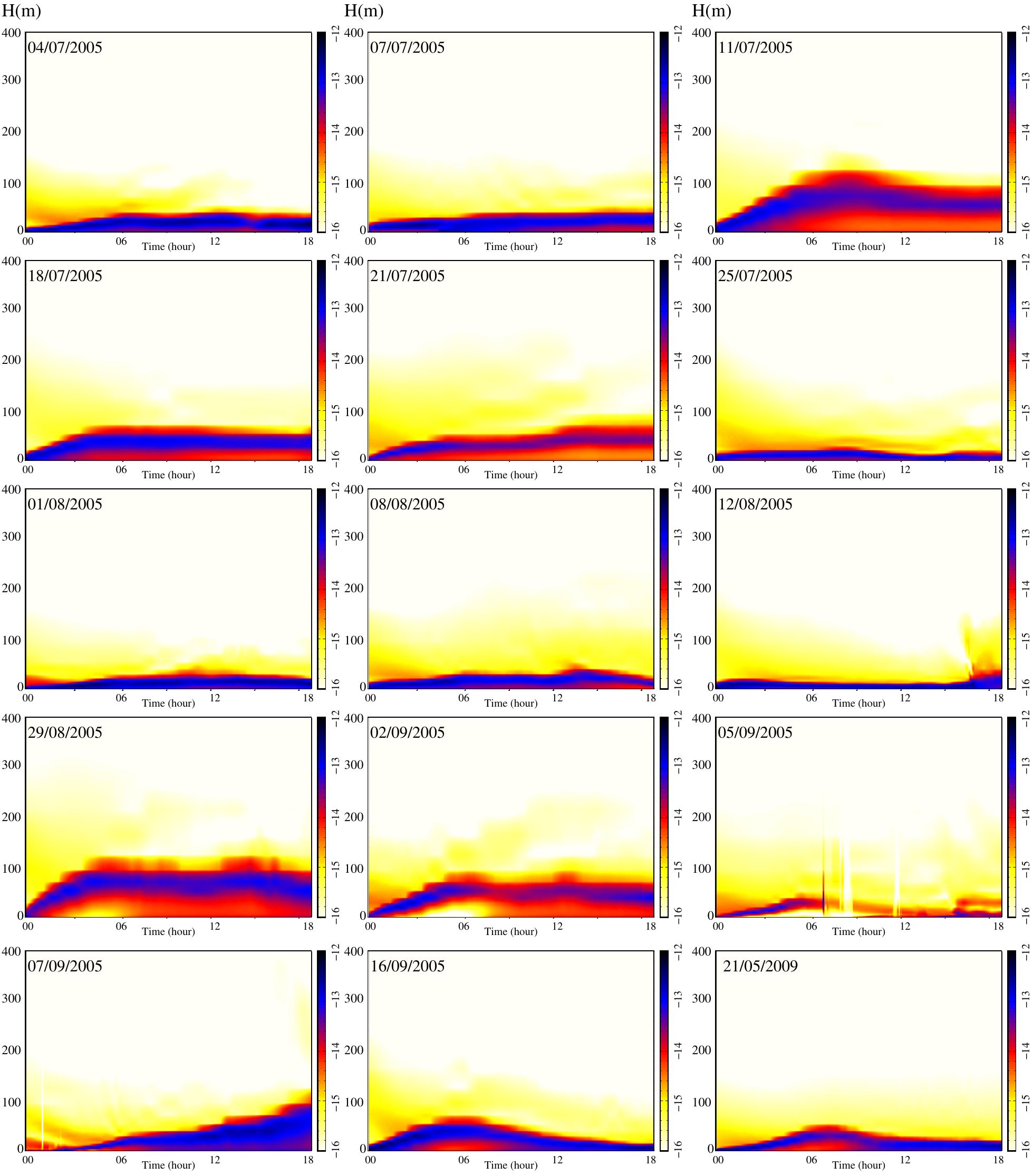}
\end{center}
\caption{Same than Figure \ref{fig1_app} but for Dome C.}
\label{fig2_app}
\end{figure*}

\begin{figure*}
\begin{center}
\includegraphics[width=\textwidth]{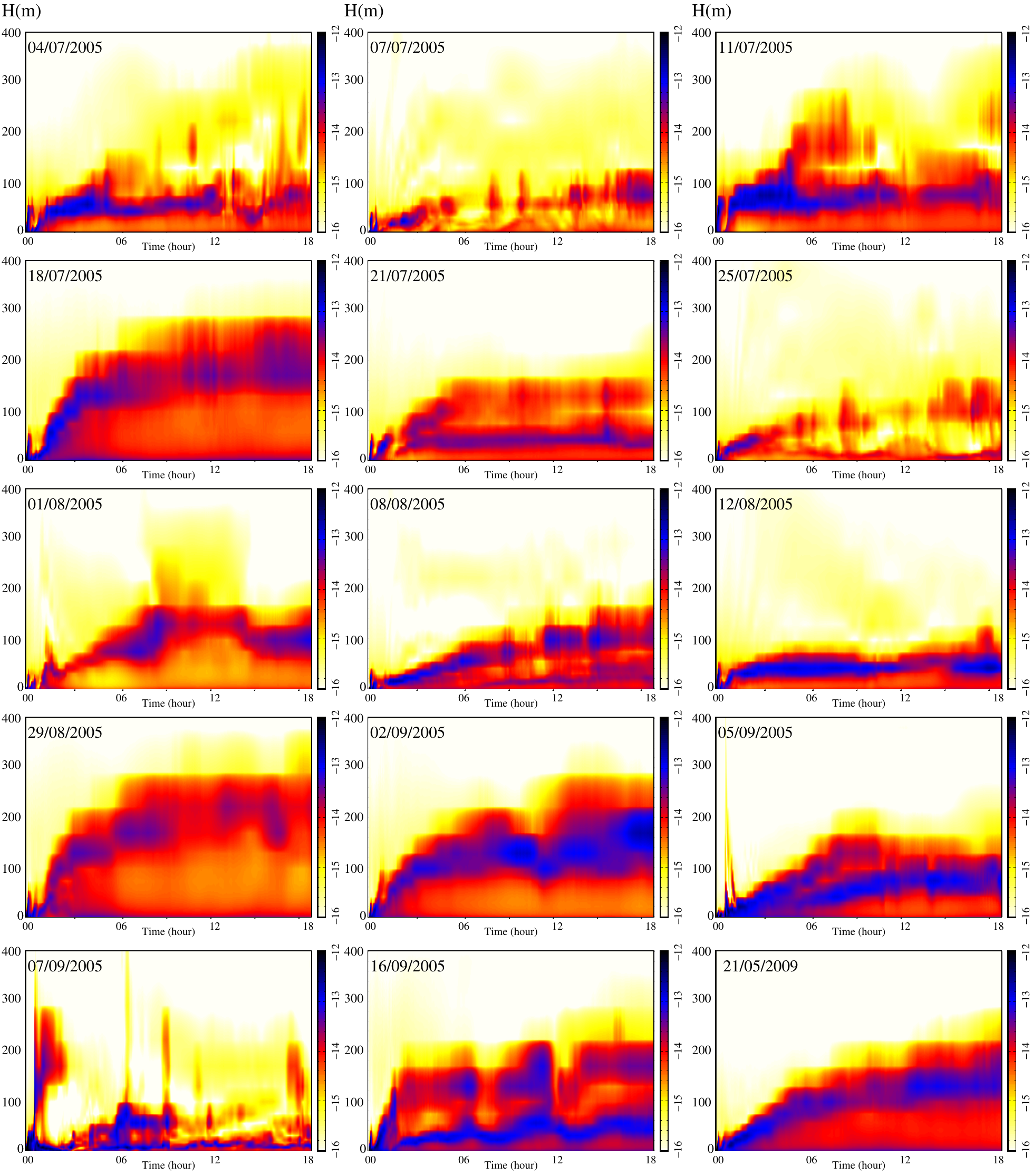}
\end{center}
\caption{Same than Figure \ref{fig1_app} but for South Pole.}
\label{fig3_app}
\end{figure*}

In this appendix we present all the individual figures of the 18-hours temporal evolution of the \CN2 for every night and at the three 
antarctic sites considered in this study (Figure \ref{fig1_app}: Dome A, Figure \ref{fig2_app}: Dome C and Figure \ref{fig3_app}: 
South Pole). 
The first couple of hours can be considered as spurious values because of the model adaptation to the ground.

\newpage
\newpage


\label{lastpage}

\begin{thebibliography}{}

\bibitem[\protect\citeauthoryear{}{1976}]{am}
Arakawa, A. \& Messinger, F., 1976, GARP Tech. Rep., 17, WMO/ICSU, Geneva,
Switzerland
\bibitem[\protect\citeauthoryear{Aristidi et al.}{2005}]{ar}
Aristidi, E., Agabi, K., Fossat, E., Azouit, M., Martin, F., Sadibekova, T., Travouillon, T., Vernin, J., Ziad, A., 2005, A\&A, 444, 651
\bibitem[\protect\citeauthoryear{Aristidi et al.}{2009}]{ar2009}
Aristidi, E., Fossat, E., Agabi, K., M\'ekarnia, D., Jeanneaux, F., Bondoux, E., Challita, Z., Ziad, A., Vernin, J., Trinquet, H., 2009, A\&A, 499, 955
\bibitem[\protect\citeauthoryear{Ashley et al.}{2010}]{as10}
Ashley, M. et al., 2010, EAS Publications Series, 40, 79
\bibitem[\protect\citeauthoryear{Asselin}{1972}]{as}
Asselin, R., 1972, Mon. Weather. Rev., 100, 487
\bibitem[\protect\citeauthoryear{Bougeault, P. \& Lacarr\`ere}{1989}]{bl}
Bougeault, P. \& Lacarr\`ere, P., 1989, Mon. Weather. Rev., 117, 1972
\bibitem[\protect\citeauthoryear{Cuxart et al.}{2000}]{cux}
Cuxart, J., Bougeault, P. and Redelsperger, J.-L.,  Q. J. R. Meteorol. Soc.,
126, 1, 2000
\bibitem[\protect\citeauthoryear{}{1975}]{gcs}
Gal-Chen, T. \& Sommerville, C. J., 1975, J. Comput. Phys., 17, 209
\bibitem[\protect\citeauthoryear{Hagelin et al.}{2008}]{ha}
Hagelin, S., Masciadri, E., Lascaux F. and Stoesz, J., 2008, MNRAS, 387, 1499
\bibitem[\protect\citeauthoryear{Lafore et al.}{1998}]{laf}
Lafore, J.-P. et al., 1998, Annales Geophysicae, 16, 90
\bibitem[\protect\citeauthoryear{Lascaux et al.}{2009}]{lf09}
Lascaux, F., Masciadri, E., Stoesz, J., Hagelin, S., 2009, MNRAS, 398, 1093
\bibitem[\protect\citeauthoryear{Lascaux et al.}{2010a}]{lf10a}
Lascaux, F., Masciadri, E., Hagelin, S., 2010a, MNRAS, 403, 1714
\bibitem[\protect\citeauthoryear{Lascaux et al.}{2010b}]{lf10b}
Lascaux, F., Masciadri, E., Hagelin, S., 2010b,
Ground-based and Airborne Telescopes III. Edited by Stepp, Larry M.; Gilmozzi, Roberto; Hall, Helen J. Proceedings of the SPIE, Volume 7733, pp. 77334E-77334E-8
\bibitem[\protect\citeauthoryear{Lawrence et al.}{2004}]{la}
Lawrence, J., Ashley, M., Tokovinin, A., Travouillon, T., 2004, Nature,
431, 278
\bibitem[\protect\citeauthoryear{Le Moigne et al.}{2009}]{lem}
Le Moigne, P., Noilhan, J., Masciadri, E., Lascaux, F., Pietroni, I., 2009, Masciadri, E. \& Sarazin, M., eds, 
Optical Turbulence - Astronomy meets Meteorology. Imperial College Press, London, p.165
\bibitem[\protect\citeauthoryear{Le Moigne et al.}{2010}]{lem10}
Le Moigne, P., Noilhan, J., Masciadri, E., Lascaux, F., Pietroni, I., 2010, Journal of Geophysical Research, submitted
\bibitem[\protect\citeauthoryear{Liu et al.}{2001}]{liu}
Liu, H., Jezek, K., Li, B., Zhao, Z., 2001, Digital media, National Snow and Ice Data Center, Boulder, CO, USA
\bibitem[\protect\citeauthoryear{Marks et al.}{1999}]{ma3}
Marks, R.D., Vernin, J., Azouit M., Manigault J.F., Clevelin C., 1999,
A\&AS, 134, 161
\bibitem[\protect\citeauthoryear{Marks et al.}{1996}]{ma2}
Marks, R.D., Vernin, J., Azouit, M., Briggs, J.W., Burton, M.G., Ashley, M.C.B., Manigault, J.F., 1996, A\&AS, 118, 385
\bibitem[\protect\citeauthoryear{Masciadri et al.}{1999a}]{m1}
Masciadri, E., Vernin, J., Bougeault, P., 1999a,
A\&ASS, 137, 185
\bibitem[\protect\citeauthoryear{Masciadri et al.}{1999b}]{m2}
Masciadri, E., Vernin, J., Bougeault, P., 1999b,
A\&ASS, 137, 203
\bibitem[\protect\citeauthoryear{masciadrietgarfias}{2001}]{masciadriegarfias01}
Masciadri, E. \& Garfias, T., 2001, 
A\&A, 366, 708
\bibitem[\protect\citeauthoryear{masciadrietal}{2001}]{masc_et_al2001}
Masciadri, E., Vernin, J., Bougeault, P., 2001,
A\&A, 365, 699
\bibitem[\protect\citeauthoryear{Masciadri and Jabouille}{2001}]{m5}
Masciadri, E. \& Jabouille, P., 2001, A\&A, 376, 727
\bibitem[\protect\citeauthoryear{Masciadri et al.}{2004}]{m3}
Masciadri, E., Avila, R., Sanchez, L. J., 2004, RMxAA, 40, 3
\bibitem[\protect\citeauthoryear{Masciadri and Egner}{2006}]{m4}
Masciadri, E. \& Egner, S., 2006, PASP, 118, 849, 1604
\bibitem[\protect\citeauthoryear{Noilhan et al.}{1989}]{np}
Noilhan J. \& Planton, S., 1999, Mon. Weather. Rev., 117, 536
\bibitem[\protect\citeauthoryear{Stein et al.}{2000}]{st}
Stein, J., Richard, E., Lafore, J.-P., Pinty, J.-P., Asencio, N., Cosma, S., 2000,
Meteorol. Atmos. Phys., 72, 203
\bibitem[\protect\citeauthoryear{Swain and Gall\'ee}{2006}]{seg}
Swain, M. \& Gall\'ee, H., 2006, PASP, 118, 1190
\bibitem[\protect\citeauthoryear{Trinquet et al.}{2008}]{tr}
Trinquet, H., Agabi, K., Vernin, J., Azouit, M., Aristidi, E.,  Fossat, E., 2008, PASP,
120, 203
\end{thebibliography}
\end{document}